# Deciphering diffuse scattering with machine learning and the equivariant foundation model: The case of molten FeO.

Ganesh Sivaraman[1,2] and Chris J. Benmore[2,3]

[1] Department of Chemical & Biomolecular Engineering, University of Illinois at Urbana-Champaign,
Urbana, Illinois 61801, USA.
[2] C-STEEL Center for Steel Electrification by Electrosynthesis, Argonne National Laboratory,
Argonne, Illinois 60438, USA.
[3] X-Ray Science Division, Advanced Photon Source, Argonne National Laboratory,
Argonne, Illinois 60438, USA.

E-mail: benmore@anl.gov

**Abstract**

Bridging the gap between diffuse x-ray or neutron scattering measurements and predicted structures derived from atom-atom pair potentials in disordered materials, has been a longstanding challenge in condensed matter physics. This perspective gives a brief overview of the traditional approaches employed over the past several decades. Namely, the use of approximate interatomic pair potentials that relate 3-dimensional structural models to the measured structure factor and its' associated pair distribution function. The use of machine learned interatomic potentials has grown in the past few years, and has been particularly successful in the cases of ionic and oxide systems. Recent advances in large scale sampling, along with a direct integration of scattering measurements into the model development, has provided improved agreement between experiments and large-scale models calculated with quantum mechanical accuracy. However, details of local polyhedral bonding and connectivity in meta-stable disordered systems still require improvement. Here we leverage MACE-MP-0; a newly introduced equivariant foundation model and validate the results against high-quality experimental scattering data for the case of molten iron(II) oxide (FeO). These preliminary results suggest that the emerging foundation model has the potential to surpass the traditional limitations of classical interatomic potentials.

## 1. Understanding disordered structures.

Finding the link between the measured static structure factor from a disordered material, and the predicted structure derived from individual atom-atom interatomic pair potentials, is a long-standing problem in condensed matter physics [1, 2]. The suitability of classical interatomic potentials in accurately predicting structure are typically assessed by comparing the Sine Fourier transform of the measured diffraction data, the total pair distribution function (PDF), to the appropriately weighted sum of the partial atom-atom PDF's. This comparison is straight forward in the case of neutron scattering where the coherent scattering lengths are constant with momentum transfer, Q, but requires consideration of the Q-dependent electronic form factors in the case of x-rays. Nevertheless, both neutron and x-ray diffraction data typically serve as rigorous tests of model structures predicted by simulation methods [3].

Ornstein-Zerneike have argued that the total correlation function is a direct effect of molecule 1 on molecule 2, plus an indirect effect of all other molecules [4]. The Ornstein-Zerneike equation can be extended using angular dependent pair distribution functions, from equations such as Hypernetted Chain Theory [5] or Percus Yevik theory [6] for hard spheres. Solving the angular dependent pair correlation function can be achieved by a series expansion of spherical harmonics, whereby the potential energy is the sum of pair interactions between atoms within different molecules. Alternatively, the use of approximate pair potentials in molecular dynamics or Monte Carlo simulations can offer approximate model structures that capture the underlying physics and chemical bonding in crystals, liquids and glasses.

A notoriously difficult example of a binary liquid x-ray PDF to measure experimentally as well as model, is shown in figure 1. The experimental data are those previously reported by Shi et al. [7] for molten $Fe_{49}O_{51}$ at 1673K containing 91%$Fe^{2+}$. Here the qualitative local polyhedral bonding and connectivity in molten FeO is reproduced by several classical Molecular Dynamics (MD) interatomic potentials, including Born-Mayer [8-10], Buckingham [11], Interionic [12] and Morse potentials [13]. However, none of these interatomic potentials are in quantitative agreement with predicting the dominant interactions associated between $Fe^{2+}$ and $O^{2-}$. The multivalent nature of Fe bonding depends several factors, including oxygen partial pressure, temperature and composition. Fe-O interactions are of fundamental importance in iron and steelmaking, where oxidation states and coordination numbers can have a direct effect on phase stability, melt viscosity, density and heat capacity.

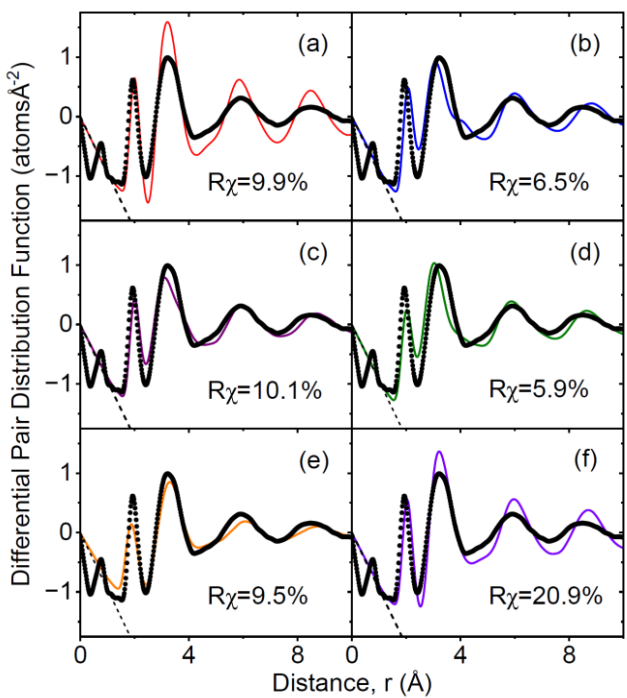

**Figure 1.** The measured x-ray differential pair distribution function, D(r), of liquid FeO (circles) compared to classical molecular dynamics simulations using six different interatomic potentials. The model partial structure factors were weighted by the Q-dependent x-ray form factors and the total S(Q) truncated at a $Q_{max}$=20Å$^{-1}$. The MD models were calculated using the potentials of (a) [9] (b) [11] (c) [12] (d) [13] (e) [10] (f) [8]. The R$\chi$-factor for each MD model is calculated with respect to the experimental data in T(r) between $r_{min}$=1Å and $R_{max}$=10Å, where a lower R$\chi$-factor indicates better agreement with experiment. Here the D(r) representation is shown rather than T(r) to highlight the structural differences.

Typically, a $\chi$-squared analysis between the computed and measured structure factor, S(Q), [14, 15] or R$\chi$-factor analysis [3, 16] between the model and experimental PDF's, is used to assess the validity of any structural model. Here the differential distribution function D(r) is related to S(Q) through a Sine Fourier transform using the Hannon-Howells-Soper formalism [17],

$$D(r) = \frac{2}{\pi}\int_0^\infty Q \frac{[S(Q)-1]}{<f(Q)>^2} sinQr \, dQ \text{ and,}$$

$$R_\chi = \left(\sum_i [T_{expt}(r_i) - T_{model}(r_i)]^2 / \sum_i T_{expt}^2(r_i)\right)^{1/2}$$

where $T(r) = D(r) + 4\pi\rho r$

where $<f(Q)>^2$ is the average x-ray scattering form factor squared and the total distribution function T(r) includes the radial bulk density term. Comparisons of the R$\chi$-factor in real space are made between the minimum and maximum atom-atom interaction distances T($r_{min}$<r<$r_{max}$). At distances shorter than $r_{min}$, oscillations in the experimental data can arise for number reasons, including systematic errors in the measurement, Fourier transform artifacts due to the finite maximum Q, and in the case of x-ray diffraction, approximations in the electron cloud distribution used to analyze the scattering data. Nonetheless these unphysical oscillations below $r_{min}$ should oscillate about the density line $-4\pi\rho r$ (see figure 1). For these reasons, Reverse Monte Carlo fits are typically refined against the S(Q) reciprocal space data [18], which has the added benefit of a straight forward comparison with statistical experimental errors.

## 2. Standard methods for modelling scattering data.

The development of parametrized classical interatomic potentials routinely used in Molecular Dynamics (MD) and Monte Carlo simulations are often based on crystalline properties, lattices and bond lengths. Popular potential types include Lennard-Jones [19], Stillinger-Weber [20], embedded-atom method [21], CHARMM [22] and AMBER [23] force fields, among many others. Although the large system sizes of these simulations enable the calculation of a systems bulk properties, a main aim of fitting potential functions is to make the potential transferable, such that it can be used to describe a different materials' structure and properties other than the one it was designed for. However, there is an inconsistency in the use of the effective pair potential approximation, since it may not effectively capture quantum mechanical phenomenon such as polarizability or many body effects, etc. While pair-theory or fixed functional form approximations greatly simplify and speed-up the computational calculations, such quantum effects are typically accounted for by using approximations to create effective pair potentials. For more accurate quantum mechanical modeling, and the prediction of both the electronic and atomic structures, the interactions between electrons needs to be considered explicitly. Density functional theory (DFT) with systematic advancements in the exchange-correlation approximations [24] has provided an effective approach for quantum mechanical treatment of materials. DFT based MD simulations have played an important in accurate modeling of condensed phases [25]. However, even the DFT based MD simulations are often limited to short simulation times (~ 10's of ps), and only a small number of atoms (100's of atoms) can be treated.

By approaching the problem from a tangential direction, Monte Carlo fits can provide a 3-dimensional structures perfectly consistent with the measured data. Methods such as Reverse Monte Carlo (RMC) [18] and Empirical Potential Structure Refinement (EPSR) [15] iteratively move atoms and/or molecules, with a given probability to avoid local minima, that ultimately improve agreement with the scattering pattern. However, these methods often lack knowledge of the underlying physics, restricting them to structure and density refinements alone. These Monte Carlo approaches are also subject to ambiguities caused by uniqueness, whereby vastly different structural models can be made to adequately fit the diffraction data with the correct density [26, 27]. The empirical potentials used in EPSR in particular are a means to drive the fitting procedure based on the difference between the model and experiment, and not meaningful in the sense of predicting the thermodynamic properties or atomic dynamics of a material. Nonetheless, EPSR has inspired our current efforts described in this perspective to try and complete the link; *using modern computing tools to create and guide realistic interatomic potentials that agree with experiment while maintaining quantum mechanical level accuracy*. An overly simplistic overview of the correlation between experimental accuracy and theoretical accuracy of existing methods and the objective of developing machine learned interatomic potentials is illustrated in figure 2.

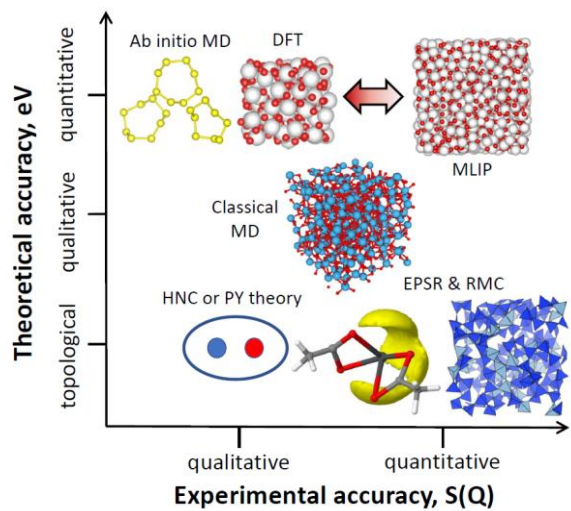

**Figure 2.** Simplistic overview of scattering model methodologies.

**3. A new perspective on understanding disorder.**

An alternative to addressing the accuracy limitations of effective pair potentials while also improving the spatio-temporal limitation of DFT-MD simulation is to directly machine learn the potential energy surfaces from *ab initio* datasets. Unlike the empirical interatomic potentials, these machines learning interatomic potentials are inherently high dimensional, which allows them to achieve accuracy as afforded by their training reference method. Many machine learning interatomic potential (MLIP) architectures have been developed since the past decade, with some of the most relevant being the Behler-Parrinello neural network [28], the Gaussian approximation potential (GAP) [29], the Spectral Neighbor Analysis Potential (SNAP) [30], the moment tensor potential (MTP) [31], ANI [32, 33], SchNet [34], DeepMD [35], NequIP [36], MACE [37] and Allegro [38]. Furthermore, addition of long-range interactions has also been proposed in recent studies using MLIP's, which would allow them to model multiply charged systems or interfaces, etc. [39, 40].

The early development of the MLIP's (except for ANI) were predominantly focused on simulations of specific chemical systems or materials under different conditions. A bottleneck to training these MLIP's was related to efficient sampling of the training configurations, as labelling these configurations still required expensive DFT calculations. To solve this challenge, active learning strategies were devised to enable efficient sampling of the most relevant minimum training configurations that would lead to optimal coverage of MLIP for the chemical space of interest. Active learning methods were devised [41] to construct quantum mechanically accurate and transferable machine learning-based models of the potential energy surface for the molecular modelling of materials. The method consists of three main steps: phase exploration, then query a small subset of that explored region for labelling, and training of the interatomic potential. These active learners have been shown to dramatically lower the number of labels needs to train the MLIPs by improving the diversity of training space. Machine learning interatomic potentials (MLIP) can maintain near *ab initio* accuracy, and enable larger system sizes through linear scaling, together with time scales comparable to classical interatomic potentials. MLIP's based on the Gaussian Approximation Potential (GAP) have been successfully trained to model liquids [42], crystals [43], defects [44], amorphous [45], multi-component materials [46], and molecules [47].

Sivaraman et al. [48] have extended the MLIP methodology by using experimental scattering data measured over a wide range of phase space to drive the procedure. An active learning scheme, initialized by model structures, iteratively improves a MLIP until agreement is obtained with experiment within a specified quantum mechanical accuracy (see next section). Using the experimental data to guide the DFT calculations is particularly important for metastable materials such as supercooled, liquid, amorphous or glassy materials, where the system is not in the lowest energy state. A critical pre-requisite of this approach is that the scattering data are of the highest quality. The appropriate experimental corrections and normalization procedures have been well documented for both neutron [49, 50] and x-ray diffraction [51, 52]. However, disordered materials, unlike crystals, do not have a periodic lattice to verify the underlying structural order. Rather, consistency checks are used verify the accuracy of the data, including that the low-r data in the PDF oscillates around the density line and the (dis-)agreement with the measured $S(Q)$ and the Fourier back-transform gives an assessment of the level of systematic error [49, 52]. A primary goal of this methodology is to develop interatomic potentials that encompass all condensed matter phases, both solid and liquid, that are transferrable. The workflow for mapping MLIP's is shown in figure 3.

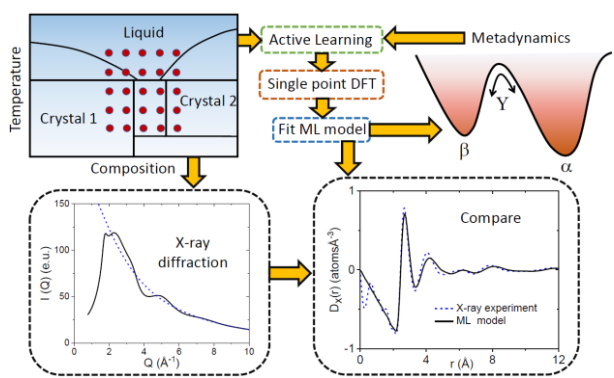

**Figure 3.** Workflow for mapping machine learned interatomic potentials across phase space. (a) Sample the configuration space. (b) Perform single point DFT for the AL samples and fit the ML model. (c) Enrich the configuration space by using meta-dynamics on the ML based MD. (d) Diffraction experiments (e) Perform rigorous validation of ML driven MD simulation using pair distribution functions.

## 4. Experimentally driven MLIP's.

Experimentally, carefully analysed x-ray and neutron diffraction data can provide rigorous benchmarks for discerning between different structural models of disordered materials. Progress toward the automated fitting of MLIP's using experimental PDFs and active learning have been made in this regard using two approaches. Firstly, reproducing the high temperature phases of the refractory oxide $HfO_2$ [48], has been made using an automated closed loop via an active learner. Crystal phases based on x-ray and neutron diffraction data are used to initialize and sequentially improve an unsupervised machine-learning model over a predetermined range of phase space. The resulting MD simulations were able to reproduce all the experimental phases with near *ab initio* precision [48]. Particularly, the experiment driven MLIP work provided clarity on the ambiguous high-temperature phase behaviour of hafnia immediately preceding melting, demonstrating agreement between simulations and experiments on the mixed cubic and tetragonal phases near the melting point. In the second approach, the initial structure of ionic liquids has been created using classical MD simulations, and down-sampled by using an active learning algorithm. Subsequently, iterative DFT calculations are performed, and a MLIP potential developed until a specified energy threshold is reached. Atomistic models developed using this method have been applied to molten LiCl [53], NaCl [44] and LiCl-KCl [54]. Such approaches provide new physical insights into the temperature-dependent coordination environment of liquids, together with property information including density, self-diffusion constants, thermal conductivity, and ionic conductivity [55].

In the example of liquid NaCl, a GAP model developed by Tovey et al. [44] was trained with 1000 atomic configurations and obtained a DFT accuracy of within 1.5 meV/atom. The MLIP enabled the analysis of the high-temperature molten salt properties on large systems (~10,000 atoms) and longer time scales (>1 ns), currently inaccessible to *ab initio* simulations. Here, the GAP model reciprocal and real space functions are compared to subsequently measured partial $S(Q)$ and PDF's obtained from Neutron Diffraction Isotopic Substitution (NDIS) measurements, essentially a double difference method [56], see figure 4. An R$\chi$-factor analysis yields 3-4% at the partial PDF level, well within the experimental errors, and a considerably more rigorous test of any model than that provided by the total neutron or x-ray pair distribution functions. In addition, the GAP model is in very good agreement with previously published experimental diffusion constants, with a root-mean-square deviation of <$0.05\times10^{-8}$ $m^2s^{-1}$ [44]. Consequently, these results demonstrate that GAP models are able to accurately capture the many-body interactions necessary to model both the structure and dynamics ionic systems.

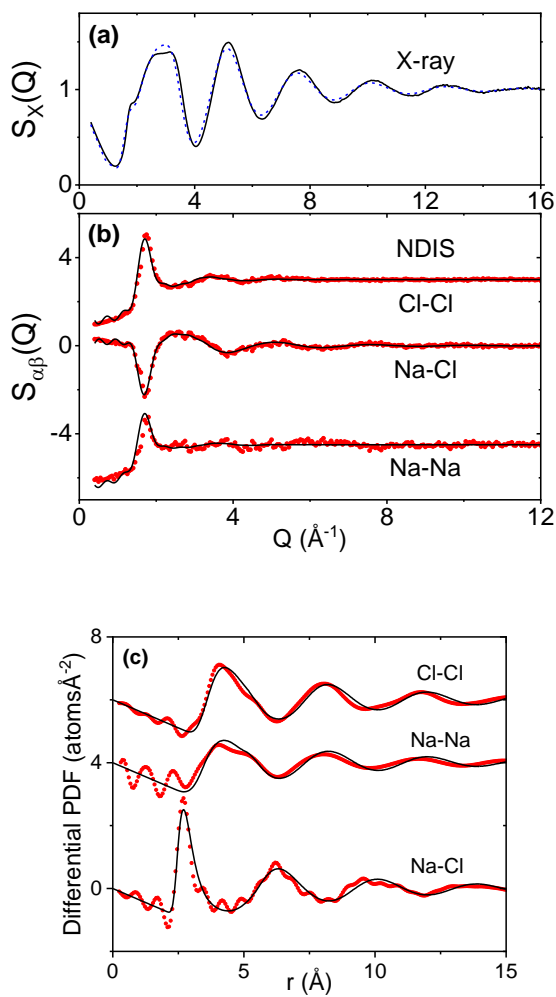

**Figure 4.** Liquid NaCl GAP model compared to the measured (a) x-ray structure factor and (b) partial pair structure factors determined from neutron diffraction isotopic substitution (NDIS) [56] (c) associated partial differential pair distribution functions and their associated R$\chi$-factors calculated between $r_{min}$=2Å and $r_{max}$=10Å.

However, limitations in DFT reference dataset or the approximation can lead to MLIP's to predict overstructured liquid structures [57], leading to a deviation with respected to experimentally measured melt structures. Matin et al. [58] introduced a novel method to refine machine learning potentials by incorporating experimental observations, specifically focusing on the melt phase of pure aluminium. This method leverages iterative Boltzmann inversion, allowing for the integration of experimental radial distribution function data at specific temperatures to act as a correction to DFT trained MLIP's. The addition of this pair correction fixed over-structuring issues seen in melt structures predicted by MLIP's without correction. Their results [58] also showed an enhancement in the prediction of diffusion constants with the added correction. However, the authors left the fitting of RDF's from different temperatures simultaneously for providing continuous corrections to DFT-trained MLIPs for future work.

*4.1 High performance computing workflow for MLIP fitting over combinatorial chemical spaces.*

The discussion so far has centred on the development and fitting of MLIPs for specific chemical systems. We now shift our focus to recent advancements in modelling or the rapid generation of MLIPs for arbitrary chemical systems, with a particular emphasis on disordered melts. In recent work, Guo et al. [59] introduced a high-performance active learning workflow termed AL4GAP. This workflow is designed to generate compositionally transferable MLIPs over charge-neutral mixtures of arbitrary molten mixtures, spanning 11 cations (Li, Na, K, Rb, Cs, Mg, Ca, Sr, Ba, Nd, and Th) and 4 anions (F, Cl, Br, I). The authors demonstrate the efficacy of this workflow by generating and validating GAP-based MLIP models with DFT-SCAN level accuracy for five systems of increasing complexity: LiCl–KCl, NaCl–CaCl$_2$, KCl–NdCl$_3$, CaCl$_2$–NdCl$_3$, and KCl–ThCl$_4$. These high-temperature melts pose significant characterization challenges due to corrosion and radiation, complex sample environments. The emergence of such a powerful framework will aid challenging x-ray synchrotron and neutron experiments in focusing their resources on the most promising melt compositions and conditions, informed by feedback from MLIP-driven MD simulations.

**5. Emergence of foundational models in materials and chemistry.**

The materials project based [60], encompassing ~1.5 million configurations across 89 chemical elements, has set the stage for the next generation of foundational models in materials science, including M3GNet [61], CHGNet [62], and MACE-MP-0 [63]. These foundation models hold the promise of development of single large models that can learn from increasingly large chemical spaces and elements. Notably, the MACE-MP-0 model demonstrated its versatility in modelling across various chemistries in solid, liquid, and gaseous states.

In the Figure 1, we showed that FeO melt in a challenging system for many classical interatomic potentials. Here we have performed the original simulation for melting of 800 atoms of FeO using MACE-MP-0 foundation model using [64]. The starting structure is created using PACKMOL [65] consisting of 100 atoms of FeO using the starting density consistent with the experimental x-ray PDF [7]. The starting structure is melted at 2500K and cooled to a target temperature of 2000K. The final structure is replicated to 800 atoms and simulations are performed in NVT ensemble. The simulations are performed using atomic simulation environment [66]. The production simulations are performed for 150ps and the PDF's are computed from the last 125 ps. Additionally, we have also computed the PDF's using the MACE-MP-0 model with dispersion correction (labelled as MACE-MP-0+D3) [67]. This simulation of molten FeO was not aimed at comprehensiveness, but to offer insights into emerging opportunities such as foundation model through original simulation results. The results are visualized in Figure. 5. The $R\chi$-factor analysis show that MACE-MP-0 simulation shows an improved agreement with experiment in comparison to every other empirical forcefield reported in Figure 1. We should note that the simulations performed in the figure 1 using the empirical forcefield were carried out using a large simulation cell of ~6000 atoms and we would presume that a larger simulation cell could only further improve the agreement for MACE-MP-0 predicted structure with experiments. In addition, it was observed that inclusion of dispersion interactions did not lead to any improvement in the predicted structure.

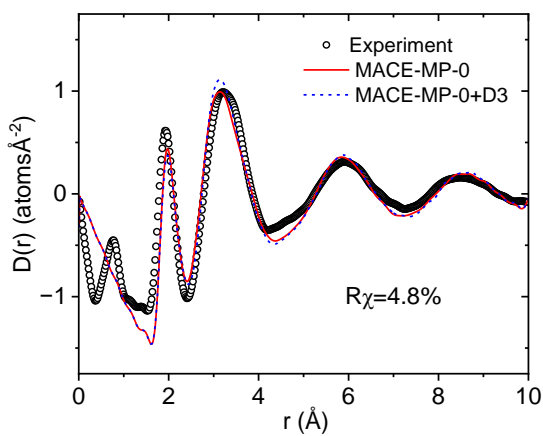

**Figure 5.** The measured x-ray differential pair distribution function, D(r), of liquid FeO at 1673K (circles) compared to the MACE-MP-0+D3 model with dispersion interactions (dashed blue line) and MACE-MP without (solid red line) at 2000K.

## 6. MLIP prospects and difficulties.

The development of accurate transferrable interatomic potentials has been a longstanding challenge in condensed matter physics. Using scattering experiments to help construct interatomic potentials is not a new idea [2], but with the advent of supercomputers, machine learning, and active learning methods, it is becoming an achievable goal. The results described in this perspective relate primarily to ionic and oxide systems, but they already indicate that MLIP models can capture the many-body interactions required to develop large scale models with quantum mechanical accuracy. Already, experimentally driven active learning methods in the field of molten salts can significantly lower the barrier to the understanding and design of materials across the periodic table. Similar active learning approaches using MLIP's applied to more complex and multi-component materials are expected to uncover new physics, particularly in disordered phases. In this regard, we argue that using experimental scattering data to drive the MLIP testing and training is essential, as the system under investigation may not necessarily be in the lowest energy state.

By scaling up the number of atoms in the simulation box, not only can material properties be more accurately predicted, but much slower quench rates of the system can also be attained. This is particularly useful in the study of metastable states such as supercooled liquids and glasses. Nonetheless, specific challenges arise when modelling metastable amorphous and glassy forms, that can cause the active learning scheme to fail. The limited sampling of configuration space in high dimensionality systems with significant free energy barriers could inhibit access to the necessary atomic arrangements. Initial attempts to widen range of possible configurations using meta-dynamics approaches [68] in the case of amorphous hafnia have not proved successful [69]. However, such schemes to increase the diversity of the local chemical environments on which the model is trained is the likely key to future progress.

For stable systems, based on the promising results for liquid FeO shown here, combining a AL4GAP type approach (to sample a vast range of targeted but diverse chemical space) to fine tune the foundation model at a higher accurate density functional theory approximation, along with direct infusion of input from experimental scattering data is a powerful tool. Moreover, this methodology could pave the way in understanding the underlying physics of structure-property relations in a multiplicity of disordered materials.

## Acknowledgements

This work was supported as part of the Center for Steel Electrification by Electrosynthesis (C-STEEL), an Energy Earthshot Research Center funded by the U.S. Department of Energy, Office of Science, Basic Energy Sciences (BES) and Advanced Scientific Computing Research (ASCR). This research used resources of the Advanced Photon Source, a U.S. Department of Energy (DOE) Office of Science user facility at Argonne National Laboratory and is based on research supported by the U.S. DOE Office of Science-Basic Energy Sciences, under contract No. DE-AC02-06CH11357. This work utilizes resources supported by the National Science Foundation's Major Research Instrumentation program, grant No. 1725729, as well as the University of Illinois at Urbana-Champaign [63].

**References**

[1]  Allen M P and Tildesley D J 2017 *Computer Simulation of Liquids*: University Scholarship Online)

[2]  Egelstaff P A 1994 *An Introduction to the Liquid State*: Oxford University Press)

[3]  Wright A C 1994 Neutron scattering from vitreous silica. V. The structure of vitreous silica: What have we learned from 60 years of diffraction studies? *Journal of Non-Crystalline Solids* **179** 84-115

[4]  Ornstein L S and Zernike F 1914 Accidental deviations of density and opalescence at the critical point of a single substance *Proc. Acad. Sci. Amsterdam* **17** 793-806

[5]  van Leeuwen J M J, Groeneveld J and de Boer J 1959 New method for the calculation of the pair correlation function. I *Physica* **25** 792-808

[6]  Percus J K and Yevick G J 1958 Analysis of Classical Statistical Mechanics by Means of Collective Coordinates *Physical Review* **110** 1-13

[7]  Shi C, Alderman O L G, Tamalonis A, Weber R, You J and Benmore C J 2020 Redox-structure dependence of molten iron oxides *Communications Materials* **1** 80

[8]  Belashchenko D K 1996 Computer simulation of non-crystalline oxides MeO and Me$_2$O$_3$ *Journal of Non-Crystalline Solids* **205-207** 212-5

[9]  Rossano S, Ramos A Y and Delaye J M 2000 Environment of ferrous iron in CaFeSi$_2$O$_6$ glass; contributions of EXAFS and molecular dynamics *Journal of Non-Crystalline Solids* **273** 48-52

[10] Seo W-G and Tsukihashi F 2005 Thermodynamic and Structural Properties for the FeO$_n$-SiO$_2$ System by using Molecular Dynamics Calculation *Materials Transactions* **46** 1240-7

[11] Yang Z, Wang B and Cormack A N 2016 The local structure of Fe in Li(Al, Fe)Si$_2$O$_6$ glasses from molecular dynamics simulations *Journal of Non-Crystalline Solids* **444** 16-22

[12] Guillot B and Sator N 2007 A computer simulation study of natural silicate melts. Part I: Low pressure properties *Geochimica et Cosmochimica Acta* **71** 1249-65

[13] Pedone A, Malavasi G, Menziani M C, Cormack A N and Segre U 2006 A New Self-Consistent Empirical Interatomic Potential Model for Oxides, Silicates, and Silica-Based Glasses *The Journal of Physical Chemistry B* **110** 11780-95

[14] McGreevy R L 2001 Reverse Monte Carlo modelling *Journal of Physics: Condensed Matter* **13** R877

[15] Soper A K 2007 Joint structure refinement of x-ray and neutron diffraction data on disordered materials: application to liquid water *Journal of Physics: Condensed Matter* **19** 335206

[16] Toby B H 2006 R factors in Rietveld analysis: How good is good enough? *Powder Diffraction* **21** 67-70

[17] Keen D 2001 A comparison of various commonly used correlation functions for describing total scattering *Journal of Applied Crystallography* **34** 172-7

[18] McGreevy R L and Pusztai L 1988 Reverse Monte Carlo Simulation: A New Technique for the Determination of Disordered Structures *Molecular Simulation* **1** 359-67

[19] Jones J E and Chapman S 1997 On the determination of molecular fields.—I. From the variation of the viscosity of a gas with temperature *Proceedings of the Royal Society of London. Series A, Containing Papers of a Mathematical and Physical Character* **106** 441-62

[20] Stillinger F H and Weber T A 1985 Computer simulation of local order in condensed phases of silicon *Physical Review B* **31** 5262-71

[21] Daw M S and Baskes M I 1984 Embedded-atom method: Derivation and application to impurities, surfaces, and other defects in metals *Physical Review B* **29** 6443-53

[22] MacKerell A D, Jr., Bashford D, Bellott M, Dunbrack R L, Jr., Evanseck J D, Field M J, Fischer S, Gao J, Guo H, Ha S, Joseph-McCarthy D, Kuchnir L, Kuczera K, Lau F T K, Mattos C, Michnick S, Ngo T, Nguyen D T, Prodhom B, Reiher W E, Roux B, Schlenkrich M, Smith J C, Stote R, Straub J, Watanabe M, Wiórkiewicz-Kuczera J, Yin D and Karplus M 1998 All-Atom Empirical Potential for Molecular Modeling and Dynamics Studies of Proteins *The Journal of Physical Chemistry B* **102** 3586-616

[23] Wang J, Cieplak P and Kollman P A 2000 How well does a restrained electrostatic potential (RESP) model perform in calculating conformational energies of organic and biological molecules? *Journal of Computational Chemistry* **21** 1049-74

[24] Perdew J P 2013 Climbing the ladder of density functional approximations *MRS Bulletin* **38** 743-50

[25] Burke K 2012 Perspective on density functional theory *The Journal of Chemical Physics* **136** 150901

[26] Evans R 1990 Comment on Reverse Monte Carlo Simulation *Molecular Simulation* **4** 409-11

[27] Soper A K 2007 On the uniqueness of structure extracted from diffraction experiments on liquids

and glasses *Journal of Physics: Condensed Matter* **19** 415108
[28] Behler J and Parrinello M 2007 Generalized Neural-Network Representation of High-Dimensional Potential-Energy Surfaces *Physical Review Letters* **98** 146401
[29] Bartók A P, Payne M C, Kondor R and Csányi G 2010 Gaussian Approximation Potentials: The Accuracy of Quantum Mechanics, without the Electrons *Physical Review Letters* **104** 136403
[30] Thompson A P, Swiler L P, Trott C R, Foiles S M and Tucker G J 2015 Spectral neighbor analysis method for automated generation of quantum-accurate interatomic potentials *Journal of Computational Physics* **285** 316-30
[31] Shapeev A V 2016 Moment Tensor Potentials: A Class of Systematically Improvable Interatomic Potentials *Multiscale Modeling \& Simulation* **14** 1153-73
[32] Smith J S, Isayev O and Roitberg A E 2017 ANI-1: an extensible neural network potential with DFT accuracy at force field computational cost *Chemical Science* **8** 3192-203
[33] Smith J S, Nebgen B, Lubbers N, Isayev O and Roitberg A E 2018 Less is more: Sampling chemical space with active learning *The Journal of Chemical Physics* **148** 241733
[34] Schütt K T, Sauceda H E, Kindermans P J, Tkatchenko A and Müller K R 2018 SchNet – A deep learning architecture for molecules and materials *The Journal of Chemical Physics* **148** 241722
[35] Wang H, Zhang L, Han J and E W 2018 DeePMD-kit: A deep learning package for many-body potential energy representation and molecular dynamics *Computer Physics Communications* **228** 178-84
[36] Batzner S, Musaelian A, Sun L, Geiger M, Mailoa J P, Kornbluth M, Molinari N, Smidt T E and Kozinsky B 2022 E(3)-equivariant graph neural networks for data-efficient and accurate interatomic potentials *Nature Communications* **13** 2453
[37] Batatia I, Kovacs D P, Simm G, Ortner C and Csanyi G 2022 MACE: Higher Order Equivariant Message Passing Neural Networks for Fast and Accurate Force Fields. ed S Koyejo*, et al.* pp 11423--36
[38] Musaelian A, Batzner S, Johansson A, Sun L, Owen C J, Kornbluth M and Kozinsky B 2023 Learning local equivariant representations for large-scale atomistic dynamics *Nature Communications* **14** 579
[39] Ko T W, Finkler J A, Goedecker S and Behler J 2023 Accurate Fourth-Generation Machine Learning Potentials by Electrostatic Embedding *Journal of Chemical Theory and Computation* **19** 3567-79
[40] Staacke C G, Heenen H H, Scheurer C, Csányi G, Reuter K and Margraf J T 2021 On the Role of Long-Range Electrostatics in Machine-Learned Interatomic Potentials for Complex Battery Materials *ACS Applied Energy Materials* **4** 12562-9
[41] L. Zhang D-Y L, H. Wang, R. Car, and E. Weinan 2019 Active learning of uniformly accurate interatomic potentials for materials simulation *Physical Review Materials* **3** 023804
[42] Sivaraman G, Krishnamoorthy A N, Baur M, Holm C, Stan M, Csányi G, Benmore C and Vázquez-Mayagoitia Á 2020 Machine-learned interatomic potentials by active learning: amorphous and liquid hafnium dioxide *npj Computational Materials* **6** 104
[43] Bartók A P, Kermode J, Bernstein N and Csányi G 2018 Machine Learning a General-Purpose Interatomic Potential for Silicon *Physical Review X* **8** 041048
[44] S. Tovey, A.N. Krishnamoorthy, G. Sivaraman, J. Guo, C. Benmore, A. Heuer and Holm C 2020 DFT Accurate Interatomic Potential for Molten NaCl from Machine Learning *Journal of Physical Chemistry C* **124** 25760-8
[45] Deringer V L, Bernstein N, Bartók A P, Cliffe M J, Kerber R N, Marbella L E, Grey C P, Elliott S R and Csányi G 2018 Realistic Atomistic Structure of Amorphous Silicon from Machine-Learning-Driven Molecular Dynamics *The Journal of Physical Chemistry Letters* **9** 2879-85
[46] Jinnouchi R, Lahnsteiner J, Karsai F, Kresse G and Bokdam M 2019 Phase Transitions of Hybrid Perovskites Simulated by Machine-Learning Force Fields Trained on the Fly with Bayesian Inference *Physical Review Letters* **122** 225701
[47] Cole D J, Mones L and Csányi G 2020 A machine learning based intramolecular potential for a flexible organic molecule *Faraday Discussions* **224** 247-64
[48] G. Sivaraman, L. Gallington, A.N. Krishnamoorthy, M. Stan, G. Csányi, Á. Vázquez-Mayagoitia and Benmore C J 2021 Experimentally Driven Automated Machine-Learned Interatomic Potential for a Refractory Oxide *Physical Review Letters* **126** 156002
[49] Fischer H E, Barnes A C and Salmon P S 2006 Neutron and x-ray diffraction studies of liquids and glasses *Reports on Progress in Physics* **69** 233
[50] Hannon A C, Howells W S and Soper A K 1990 ATLAS : A Suite of Programs for the Analysis of Time-of-flight Neutron Diffraction Data from Liquid and Amorphous Samples.
[51] Benmore C J 2023 *Comprehensive Inorganic Chemistry III (Third Edition),* ed J Reedijk and K R Poeppelmeier (Oxford: Elsevier) pp 384-424
[52] Gallington L C, Wilke S K, Kohara S and Benmore C J 2023 Review of Current Software for Analyzing Total X-ray Scattering Data from Liquids. In: *Quantum Beam Science,*
[53] G. Sivaraman, J. Guo, L. Ward, N. Hoyt, M. Williamson, I. Foster, C. Benmore and Jackson N 2021 Automated Development of Molten Salt

[54] Guo J, Ward L, Babuji Y, Hoyt N, Williamson M, Foster I, Jackson N, Benmore C and Sivaraman G 2022 A Composition-Transferable Machine Learning Potential for LiCl-KCl Molten Salts Validated by HEXRD. American Chemical Society (ACS))

Machine Learning Potentials: Application to LiCl *Journal of Physical Chemistry Letters* 4278

[55] Zhang J, Pagotto J, Gould T, Duignan, T 2023 Accurate, fast and generalisable first principles simulation of aqueous lithium chloride *arXiv [physics.chem-ph]* 2310.12535

[56] Zeidler A, Salmon P S, Usuki T, Kohara S, Fischer H E and Wilson M 2022 Structure of molten NaCl and the decay of the pair-correlations *The Journal of Chemical Physics* **157** 094504

[57] Smith J S, Nebgen B, Mathew N, Chen J, Lubbers N, Burakovsky L, Tretiak S, Nam H A, Germann T, Fensin S and Barros K 2021 Automated discovery of a robust interatomic potential for aluminum *Nature Communications* **12** 1257

[58] Matin S, Allen A E A, Smith J, Lubbers N, Jadrich R B, Messerly R, Nebgen B, Li Y W, Tretiak S and Barros K 2024 Machine Learning Potentials with the Iterative Boltzmann Inversion: Training to Experiment *Journal of Chemical Theory and Computation* **20** 1274-81

[59] Guo J, Woo V, Andersson D A, Hoyt N, Williamson M, Foster I, Benmore C, Jackson N E and Sivaraman G 2023 AL4GAP: Active learning workflow for generating DFT-SCAN accurate machine-learning potentials for combinatorial molten salt mixtures *The Journal of Chemical Physics* **159** 024802

[60] MPtrj https://figshare.com/articles/dataset/Materials_Project_Trjectory_MPtrj_Dataset/23713842.

[61] Chen C and Ong S P 2022 A universal graph deep learning interatomic potential for the periodic table *Nature Computational Science* **2** 718-28

[62] Deng B, Zhong P, Jun K, Riebesell J, Han K, Bartel C J and Ceder G 2023 CHGNet as a pretrained universal neural network potential for charge-informed atomistic modelling *Nature Machine Intelligence* **5** 1031-41

[63] Batatia I, Benner P, Chiang Y, Elena A M, Kovács D P, Riebesell J, Advincula X R, Asta M, Baldwin W J, Bernstein N, Bhowmik A, Blau S M, Cărare V, Darby J P, De S, Pia F D, Deringer V L, Elijošius R, El-Machachi Z, Fako E, Ferrari A C, Genreith-Schriever A, George J, Goodall R E A, Grey C P, Han S, Handley W, Heenen H H, Hermansson K, Holm C, Jaafar J, Hofmann S, Jakob K S, Jung H, Kapil V, Kaplan A D, Karimitari N, Kroupa N, Kullgren J, Kuner M C, Kuryla D, Liepuoniute G, Margraf J T, Magdău I-B, Michaelides A, Moore J H, Naik A A, Niblett S P, Norwood S W, O'Neill N, Ortner C, Persson K A, Reuter K, Rosen A S, Schaaf L L, Schran C, Sivonxay E, Stenczel T K, Svahn V, Sutton C, van der Oord C, Varga-Umbrich E, Vegge T, Vondrák M, Wang Y, Witt W C, Zills F and Csányi G 2023 A foundation model for atomistic materials chemistry *arXiv [physics.chem-ph]*

[64] Kindratenko V, Mu D, Zhan Y, Maloney J, Hashemi S H, Rabe B, Xu K, Campbell R, Peng J and Gropp W 2020 HAL: Computer System for Scalable Deep Learning. In: *Practice and Experience in Advanced Research Computing*: Association for Computing Machinery) pp 41–8 , numpages = 8

[65] Martínez L, Andrade R, Birgin E G and Martínez J M 2009 PACKMOL: A package for building initial configurations for molecular dynamics simulations *Journal of Computational Chemistry* **30** 2157-64

[66] Hjorth Larsen A, Jørgen Mortensen J, Blomqvist J, Castelli I E, Christensen R, Dułak M, Friis J, Groves M N, Hammer B, Hargus C, Hermes E D, Jennings P C, Bjerre Jensen P, Kermode J, Kitchin J R, Leonhard Kolsbjerg E, Kubal J, Kaasbjerg K, Lysgaard S, Bergmann Maronsson J, Maxson T, Olsen T, Pastewka L, Peterson A, Rostgaard C, Schiøtz J, Schütt O, Strange M, Thygesen K S, Vegge T, Vilhelmsen L, Walter M, Zeng Z and Jacobsen K W 2017 The atomic simulation environment—a Python library for working with atoms *Journal of Physics: Condensed Matter* **29** 273002

[67] Grimme S, Antony J, Ehrlich S and Krieg H 2010 A consistent and accurate ab initio parametrization of density functional dispersion correction (DFT-D) for the 94 elements H-Pu *The Journal of Chemical Physics* **132** 154104

[68] Bonati L and Parrinello M 2018 Silicon Liquid Structure and Crystal Nucleation from Ab Initio Deep Metadynamics *Physical Review Letters* **121** 265701

[69] Sivaraman G, Csanyi G, Vazquez-Mayagoitia A, Foster I T, Wilke S K, Weber R and Benmore C J 2022 A Combined Machine Learning and High-Energy X-ray Diffraction Approach to Understanding Liquid and Amorphous Metal Oxides *Journal of the Physical Society of Japan* **91** 091009